\documentclass{ws-ijmpb}

\usepackage{graphicx}

\begin{document}

\markboth{T. Senthil}{Deconfined criticality}

\catchline{}{}{}{}{}

\title{Quantum matters: Physics beyond Landau's paradigms\footnote{Proceedings of conference on ``Recent Progress in Many Body Theories'', Santa Fe, August 2004}}

\author{T. Senthil}
\address{Department of Physics, Massachusetts Institute of
Technology, Cambridge, Massachusetts 02139
email: senthil@mit.edu}


\maketitle


\begin{abstract}

Central to our understanding of quantum many particle physics are two ideas due to Landau. The first is the notion of the electron as a
well-defined quasiparticle excitation in the many body state. The second is that of the order parameter to distinguish different states of matter.
Experiments in a number of correlated materials raise serious suspicions about the general validity of either notion. A growing body of theoretical work
has confirmed these suspicions, and explored physics beyond Landau's paradigms. This article provides an overview of some of these theoretical developments.

\end{abstract}



\section{Introduction}

Much of our understanding of quantum many electron physics rests on
two central ideas which may be traced to Landau. The first is the notion that the
electron retains its integrity as a quasiparticle excitation above the quantum ground state
of the many particle system. This notion underlies Landau's celebrated Fermi Liquid theory of metals but is also
shared by many familiar phases of matter (band insulators, BCS superconductors, spin density waves, .....).
This is true even if microscopically
the electrons interact reasonably strongly with each other. The other important idea of Landau is
that of the order parameter
to classify and distinguish phases of matter. Closely related is the notion of spontaneously broken symmetry -
indeed the Landau order
parameter quantifies the amount of symmetry breaking in any ordered phase. The concept of the
order parameter plays an important role in
phase transition theory. The universal critical singularities at second order phase transitions
are usually attributed to the long wavelength
fluctuations of the order parameter degrees of freedom. When combined with general renormalization
group ideas this gives a sophisticated
theoretical framework - often known as the Landau-Ginzburg-Wilson(LGW) paradigm - for describing
phase transition phenomena.

Both of these two notions - the integrity of the electron and the
Landau order parameter - are so fundamental that they are
routinely taught at the early stages of a solid state physics
education. Remarkably in the last several years, a number of
experimental developments have challenged the general
applicability of either of these two basic notions to quantum
condensed matter. The best known is the phenomenon of the quantum
Hall effect that occurs in a two dimensional electron gas in high
magnetic field\cite{fqhe}. The quantum Hall systems are  striking and
well-established examples of the violation of both of Landau's
paradigms. Indeed the electron does not in general survive as a
quasiparticle\cite{fcfqhe} in fractional quantum Hall states nor is the order
in such a state captured by a local Landau order parameter\cite{wento1}.

Since the discovery and explanation of the quantum Hall effect, a
number of other remarkable phenomena have been found in other
correlated electron systems which are poorly understood. Perhaps
the most notorious is the problem of high temperature
superconductivity but there are many others as well. Examples
include a host of ``non-fermi liquid'' phenomena in the rare-earth
intermetallics known as the heavy fermions\cite{hfexp}, various kinds of
materials near Mott metal-insulator transitions, etc. Many of
these long standing problems in condensed matter theory seem not
to yield to any conventional thinking. They presumably require new
methods of attack, new languages, and perhaps even new conceptual
advances. At any rate these experimental discoveries certainly
lead to the suspicion that Landau's paradigms may breakdown in
serious ways in correlated many electron systems. In the last few
years, this suspicion has been strikingly confirmed by a variety
of theoretical advances that unquestionably demonstrate the
inadequacies of Landau's paradigms for a general understanding of
correlated matter.

This article provides an overview of these theoretical
developments. We begin by briefly discussing phases of quantum
matter that do not fit in with Landau's ideas. We then discuss
very recent work on the breakdown of the Landau paradigm at zero
temperature `quantum' phase transitions.

\section{Breakdown of Landau paradigms in correlated quantum phases}

As mentioned above the fractional quantum Hall effect provides a
well-established (and by now old) example of a correlated phase
that violates Landau's paradigms. An even older example is
provided by one dimensional systems such as (half-integer) quantum
spin chains or polyacetylene\cite{spnchn}. A more modern realization of one
dimensional physics occurs in the Carbon nanotubes\cite{nntube}. In these
examples, the electron does not retain its integrity as a
quasiparticle excitation. Rather the excitations have quantum
numbers that are fractions of those of the electron.

One of the most interesting theoretical developments in the last
several years is the realization that such `broken electron'
phenomena are {\em not} restricted to such extreme situations as
one dimension or two dimensions in strong magnetic fields. Indeed
it has become clear that electrons can break apart in regular
solids with strong electron-electron interaction in any spatial
dimension. The electron does not survive as a quasiparticle in
such phases of matter - instead there are excitations with quantum
numbers that are fractions of those of the electron.

A great deal has been learnt about these `fractionalized' phases
in $d \geq 2$  at zero magnetic field theoretically. The structure
of the excitation spectrum (or more generally the structure of the
low energy effective field theory) has been elucidated\cite{ReSaSpN,wen91,bfn,z2long,frnyk,levwen}. These
phases have a certain kind of `order' that is not captured by a
local Landau order parameter. Rather the ordering is a global
property of the many electron ground state wavefunction\cite{rech,wen91,topsf} - often
referred to as `topological order'. This kind of order generalizes
and is indeed distinct from the old notion of spontaneously broken
symmetry. Several concrete and simple microscopic models which
display these phenomena exist in both two\cite{kit,mstri,misgkag,bsf,motse1,motse2,motz3,wenkit} and three spatial
dimensions\cite{motse2,wen3d,hermp,ms3d}. Prototypical ground state wavefunctions for a number
of such fractionalized states can be written and explicitly shown
to possess properties (such as topological order) expected on
general grounds\cite{rech,ivse}. Finally there even exist some ideas on how to
directly detect certain kinds of topological order in experiments\cite{topsf,sfjj,iof}.

All of this is spectacular theoretical progress - much of it
happened in the last five or so years and builds on important
ideas and results\cite{rvb,rech,ReSaSpN,wen91} from the early days of high-$T_c$ theory. In
particular many of the theoretical criticisms levelled against
these ideas have now been satisfactorily answered.
However there
still is no unambiguous identification of such broken electron
phenomena in experiments other than the previously established
instances (FQHE and $d = 1$).

Where else might it happen? The theoretical understanding provides
some hints. It has long been appreciated that frustrated quantum
magnets may be a good place to look for such physics. It has also
become clear that other promising candidates are not-so-strongly
correlated materials. This may be seen explicitly in some of the
microscopic models showing fractionalization where it appears in
intermediate correlation regimes where neither kinetic nor
potential energy overwhelmingly dominates the other\cite{motse2}. Further
support is provided by the observation that in spin systems
fractionalization is promoted by multi-particle ring exchange
terms\cite{lhuill,z2long,bsf}ch terms become increasingly important for the spin
physics of Mott insulators as one moves away from the very strong
interaction limit (decreasing $U$ in a Hubbard model description).
Thus Mott insulators that are not too deeply into the insulating
phase or quantum solids such as He-$3$ or He-$4$ near melting may
be good places to look as well.

\section{Breakdown of Landau paradigms at quantum phase
transitions}

We now turn to the breakdown of Landau's paradigms at zero
temperature `quantum' phase transitions. That this might happen
was originally hinted at by various distinct kinds of observations in the
literature. First as reviewed in the previous section, Landau
order parameters do not necessarily capture the true order in
quantum phases. Then it is quite natural that transitions out of
such phases are not described by Landau ideas either. For instance
continuous transitions exist between distinct quantum Hall states
which clearly cannot be described in terms of simple order
parameter fluctuations a'la Landau. But what about transitions out of
of phases in which Landau order parameters do capture the order?
Here at least one might have hoped for Landau ideas on phase
transitions to work. We now review recent work showing that even
in this case the Landau paradigm breaks down.

The possibility of such a breakdown is suggested by two different
observations. The first is in numerical calculations on various
quantum transitions that see a direct second order quantum phase
transition between two phases with different broken symmetry
characterized by two apparently independent order parameters\cite{assaad,sandvik}. This
is in general forbidden within the Landau approach to phase
transitions except at special multicritical points. A similar
phenomenon is also seen in experiments\cite{graf} on the heavy fermion
compound $UPt_3$. At low temperatures, this is a superconductor
(believed to be triplet paired). Upon doping $Pd$ into the $Pt$
site, the superconductivity very quickly disappears and is
replaced instead by an antiferromagnetic metal. Within the
resolution of the existing experiments these two different kinds
of order (superconductivity and antiferromagnetism) seem to be
separated by a direct second order transition - within Landau
order parameter theory this too would be a special accident.
However the surprising frequency with which such
`Landau-forbidden' quantum transitions show up suggests a
reexamination of the validity of the Landau paradigm itself.

A second and perhaps more important reason to suspect the general
validity of the Landau paradigm comes from a number of fascinating
experiments probing the onset of magnetic long range order in the
heavy fermion metals\cite{hfexp}. Remarkably the behavior right at the quantum
transition between the magnetic and non-magnetic metallic phases
is very strikingly different from that of a fermi liquid. The
natural assumption is to attribute the non-fermi liquid physics to
the universal critical singularities of the quantum critical
point. Within the Landau paradigm these will be due to long
wavelength long time fluctuations of the natural magnetic order
parameter. In other words the hope is that Landau's ideas on phase
transitions may perhaps be used to kill Landau's theory of Fermi
Liquids. However theories associating the critical singularities
with fluctuations of the natural magnetic order parameter in a
metallic environment\cite{hertz} seem to have a hard time explaining the
observed non-fermi liquid phenomena. This failure once again fuels
the suspicion that perhaps the Landau approach to phase
transitions is incorrect. Specifically other phenomena such as the
possible loss of Kondo screening of local moments may contribute\cite{hfth} to and perhaps even dominate
the critical singularities\cite{sesavo}. This kind of thinking - particularly the latter possibility - is clearly
outside the LGW framework for critical phenomena. In other words
{\em the Landau order parameter (even if present) may distract from
the fluctuations responsible for the true critical behavior}.

These suspicions have been strikingly confirmed in recent theoretical work\cite{deccp,dcprb} on
quantum phase transitions in insulating magnets in two spatial dimension. As usual insulating magnets provide a good
theoretical laboratory to study phase transition phenomena. A number of results have been found
which quite clearly demonstrate the failure of LGW theory at certain (but not all) quantum phase transitions.
In all the examples studied so far the critical phenomenology is instead apparently most conveniently described in
terms of objects that carry fractional quantum numbers and which interact with each other through emergent gauge forces.
These fractional objects do not necessarily exist (as good excitations) in the two phases but become
useful degrees of freedom at the quantum critical point. This kind of phenomenon has been dubbed `deconfined' quantum criticality
- with a sharp and specific meaning of the term `deconfined'.

\begin{figure}[tb]
\centering
\includegraphics[width=3.0in]{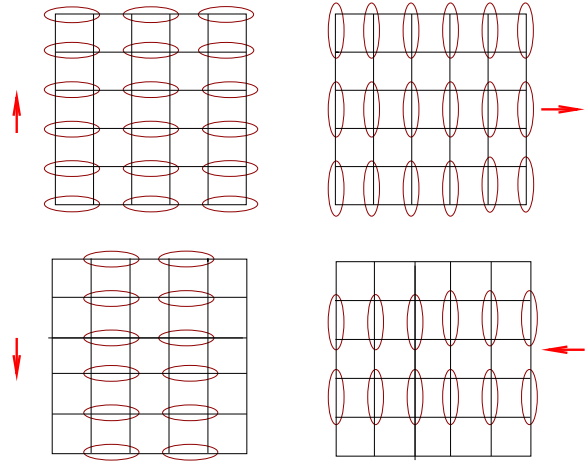}
\caption{Schematic picture of a columnar VBS state showing the four degenerate ground states.
The encircled lines
represent the bonds across which the spins are paired into a
valence bond. The four ground states are associated with four
different orientations of a $Z_4$ clock order parameter.}
\label{vbsfig}
\end{figure}

Consider a spin-$1/2$ antiferromagnet on a two dimensional square lattice described by a Hamiltonian of the general form
\begin{equation}
H = J\sum_{<rr'>} \vec S_r. \vec S_{r'} +.......
\end{equation}
The $\vec S_r$ are spin-$1/2$ operators and $J >0$ is the nearest neighbour exchange constant.
The ellipses represent other interactions such as a diagonal exchange or multiparticle ring exchange that may be tuned to drive phase transitions.
In the absence of these extra terms the ground state is known to have long ranged Neel order. The corresponding order parameter is a vector in spin space
$\vec N \sim (-1)^{(x+y)}\vec S_r$.  For suitable choices of the extra terms it is
expected that the ground state will not have long range Neel order even at zero temperature. The simplest of such `quantum paramagnets' are states
known as valence bond solids(VBS) - see Fig. \ref{vbsfig} In a cartoon of such states each spin forms a singlet valence bond with one of its neighbours. The resulting dimers stack up in some
particular pattern in the VBS ground state. The resulting state clearly has spin rotation symmetry but the pattern of dimer ordering
breaks various lattice symmetries. Clearly the order parameter for the VBS state is a spin singlet that transforms non-trivially under the lattice space group operations
- it is readily constructed out of the bond energy operators $\vec S_r.\vec S_{r'}$.
  The elementary spin-carrying excitations in this phase are gapped spin triplet particles.

In a naive Landau description of the two phases that focuses only on the low energy order parameters, a
direct second order transition is not
expected except at fine tuned multicritical points. However this naive expectation has been
argued to be incorrect. A generic second order
transition is possible between these two phases with different
broken symmetries.
 The resulting critical theory is however
unusual and {\em not} naturally described in terms of the order
parameter fields of either phase.

The key reason behind this violation of naive Landauesque expectation lies in the observation\cite{hald88,avbs} that topological defects in either
order parameter carry non-trivial quantum numbers. In particular the defects in one order parameter
transform in the same way under the microscopic symmetries as the order parameter for the other phase. Thus when the defects
in, say the Neel vector configuration, proliferate they destroy long range Neel order.
However the non-trivial quantum numbers they carry induces VBS order in the resulting paramagnet\cite{ReSaSuN}.

In the theory of Ref.~\refcite{deccp,dcprb}  the natural description of the transition
is in terms of spin-$1/2$ ``spinon'' fields $z_{\alpha}$
($\alpha=1,2$ is a spinor index).  The N\'eel order parameter is
bilinear in the spinons:
\begin{equation}
  \label{eq:cp1}
  \vec{N} \sim z^\dagger \vec\sigma z^{\vphantom\dagger}.
\end{equation}
Here $\vec\sigma$ is the usual vector of Pauli matrices and
multiplication of the spinor index is implied.  The fields $z_\alpha$
create single spin-$1/2$ quanta, ``half'' that of the spin-$1$ quanta
created by the N\'eel field $\vec{N}$. The analysis of Ref. \cite{deccp,dcprb}
shows that the correct critical field theory has the action
 $\mathcal{S}_z = \int d^2 r d \tau \mathcal{L}_z$, and
\begin{eqnarray}
\mathcal{L}_{z} &=&  \sum_{a = 1}^2 |\left(\partial_{\mu} -
ia_{\mu}\right) z_{a}|^2 + s |z|^2 +
u\left(|z|^2 \right)^2 \nonumber \\
&~&~~~~~~~ +
\kappa\left(\epsilon_{\mu\nu\kappa}\partial_{\nu}a_{\kappa}\right)^2,
\label{sz}
\end{eqnarray}

This is distinct from both the $O(3)$ universality class in $D = 3$ (as might have been expected based on the Neel order parameter) or the
$Z_4$ universality class (as expected from the $Z_4$ VBS order parameter). The distinction with the $O(3)$ universality class may be a bit puzzling to
field theorists familiar with the $CP^1$  description of the $O(3)$ non-linear sigma model - the crucial point is that the gauge field in Eqn. \ref{sz}
above is non-compact. As explained in Ref. \refcite{mv} with a non-compact gauge field this model does not describe the usual $O(3)$ ordering transition in
$D = 3$ - rather it describes the transition\cite{km} in $O(3)$ models wehere `hedgehog' defects have been suppressed by hand\cite{dasl}. Ref. \refcite{mv} also
contains detailed numerical calculations of critical exponents for the transition in the model Eqn. \ref{sz}. The non-compactness of the gauge field leads
to an extra emergent conservation law (conserved gauge flux) that helps give precise meaning to the notion of deconfinement at the critical point. This conservation law
emerges only at the critical point and does not obtain away from it in either phase.

The critical behavior at this transition is strikingly anamolous  - indeed it may be viewed as the moral equivalent of `non-fermi liquid' behavior in this insulating context.
For instance the magnon spectral function is extremely broad when compared to other quantum transitions - the exponent $\eta$ is estimated\cite{mv}
to be $\approx 0.6$,
bigger by an order of magnitude as compared to the conventional $O(3)$ fixed point. This may roughly be understood as being due to the decay of magnons into
the spinon degrees of freedom. A number of other interesting properties - such as the presence of more than one diverging length/time scale - have
also been found\cite{deccp,dcprb}.

The Neel-VBS transition is not the only example of this kind of
quantum phase transition. A number of other transitions in quantum
antiferromagnetism have been shown to have many similarities with
the phenomena described above. These include the transition from
the VBS state to a gapped `spin liquid' paramagnet\cite{dcprb} and
quantum transitions between two different patterns of VBS ordering
on certain lattices\cite{vbs,frd1} (for instance in a bilayer
honeycomb lattice). Further deconfined critical {\em phases}
described by gapless Dirac-like fermionic spin-$1/2$ objects
coupled to an emergent non-compact $U(1)$ gauge field have been
shown to exist as stable quantum phases\cite{hermu1} in two space
dimensions. Thus it appears that the phenomenon of deconfined
quantum criticality is reasonably common in two dimensional
quantum magnets.

Easy plane versions of quantum spin-$1/2$ models have also been examined and shown to have Landau-forbidden transitions and asociated deconfined
quantum critical points\cite{deccp,dcprb}. These may also be fruitfuly viewed as superfluid-insulator transitions of bosons at half-filling on the square lattice.
The case of bosons at a general commensurate filling $p/q$ has been examined recently\cite{bbbss1} - again the topological defects have been shown to carry
non-trivial quantum numbers which in turn leads to non-trivial order in the insulating phase. Direct second order Landau-forbidden transitions
seem possible for a number of special fillings.

\section{Conclusions}

The developments discussed above provide a theoretically important zeroth order answer to the basic question posed by experiments in modern
correlated electron physics: Can Landau's ideas breakdown in quantum matter more generally than in one dimension or the quantum Hall effect?
While the theoretical progress at this basic level has been dramatic, we do not at present  know what role, if any, it will play in understanding
existing experiments on materials such as the cuprates or the heavy fermions. Nevertheless
the intuition gleaned from these results will hopefully suggest ways of thinking correctly about such experimental problems.


\section*{Acknowledgements}
 It is a pleasure to
thank my many collaborators on these matters - a partial list is P. Ghaemi, M. Levin, M. Hermele, O. Motrunich, A. Vishwanath, C. Lannert,
D. Ivanov, L. Balents, S. Sachdev, Matthew Fisher, P.A. Lee, N. Nagaosa, and X.-G. Wen. This research is supported by the National Science Foundation
grant DMR-0308945,  the
NEC Corporation, the Alfred P. Sloan Foundation, and
The Research Corporation.


\end{document}